\documentclass[letterpaper,11pt]{article}
\usepackage[utf8]{inputenc} 
\usepackage[margin=1in]{geometry}
\usepackage{tikz,rotating,color,subcaption,url, comment}
\usetikzlibrary{matrix,arrows,automata,shapes,patterns}
\usepackage{amsfonts,amssymb,amsmath,amsthm,dsfont}
\usepackage{array, adjustbox,makecell}
\usepackage{multirow}
\theoremstyle{definition}

\usepackage{multicol,float}
\usepackage{lscape}

\usepackage{algorithm}
\usepackage{rounddiag}

\usepackage{technote}
\usepackage{homodel}

\tikzstyle{level 1}=[level distance=3.5cm, sibling distance=3cm]
\tikzstyle{level 2}=[level distance=4.5cm, sibling distance=2cm]
\tikzstyle{level 3}=[level distance=3.5cm, sibling distance=1cm]
\tikzstyle{bag} = [text width=6em, text centered]
\tikzstyle{end} = [circle, minimum width=0pt,fill, inner sep=0pt]

\usepackage[normalem]{ulem}

\usepackage{amsthm}
\newcommand{\toto}{xxx}

\newenvironment{proofL}{\noindent{\bf
		Proof }} {\hspace*{\fill}$\Box_{Lemma~\ref{\toto}}$\par\vspace{3mm}}

\newenvironment{proofP}{\noindent{\bf
		Proof }} {\hspace*{\fill}$\Box_{Proposition~\ref{\toto}}$\par\vspace{3mm}}

\usepackage{xargs}                      % Use more than one optional parameter in a new commands
\usepackage{xcolor}  % Coloured text etc.
\usepackage[colorinlistoftodos,prependcaption,textsize=tiny]{todonotes}

\newconstruct{\FOREACH}{\textbf{for each}}{\textbf{do}}{\ENDFOREACH}{}

\newcommand\Broadcast{\textbf{broadcast}}

\newident{noop}
\newconstruct{\UPON}{\textbf{upon}}{\textbf{do}}{\ENDUPON}{}

\newcommand\Proposal{\mathsf{PROPOSE}}

\newcommand\Precommit{\mathsf{VOTE}}

\newcommand\coord{\mathsf{proposer}}

\newconstruct{\FUNCTION}{\textbf{Function}}{\textbf{:}}{\ENDFUNCTION}{}

 \renewcommand{\note}[2][default]{\relax}

\newcommand{\tr}[1]{}
\renewcommand{\tr}[1]{#1}

\newcommand\vote{vote}

\newcommand\currHeight{k}
\newcommand\currRound{t} %Round is the number of try for a given height, what we called epoch in Tendermint
\newcommand{\ind}{\ensuremath{\mathds{1}}}

\newcommand{\true}{\ensuremath{1}}
\newcommand{\false}{\ensuremath{0}}
\newcommand{\unknown}{\ensuremath{\bot}}

\newcommand{\actionPropose}{\ensuremath{action^{\text{propose}}}}
\newcommand{\actionCheck}{\ensuremath{action^{\text{check}}}}
\newcommand{\actionSend}{\ensuremath{action^{\text{send}}}}
\newcommand{\validValue}{\textit{validValue}}
\newcommand{\createValidValue}{\texttt{createValidValue}}
\newcommand{\createInvalidValue}{\texttt{createInvalidValue}}

\newcommandx{\YAG}[2][1=]{\todo[linecolor=black,backgroundcolor=teal!25,bordercolor=teal,#1]{\textcolor{teal}{\textbf{Yackolley}}\\#2}}
\newcommandx{\ST}[2][1=]{\todo[linecolor=red,backgroundcolor=red!25,bordercolor=red,#1]{#2}}
\newcommandx{\AD}[2][1=]{\todo[linecolor=blue,backgroundcolor=blue!25,bordercolor=blue,#1]{#2}}
\newcommandx{\MP}[2][1=]{\todo[linecolor=green,backgroundcolor=green!25,bordercolor=green,#1]{#2}}

\newcommand\redout{\bgroup\markoverwith
	{\textcolor{red}{\rule[0.5ex]{2pt}{0.8pt}}}\ULon}

\newcommand \nil{\textit{nil}}

\newcommand \isProposer{\texttt{isProposer}}
\newcommand{\isValid}{\ensuremath{\texttt{isValid}}}

\newcommand \setActionCheck{\ensuremath{\sigma^{\text{check}}}}
\newcommand \setActionSend{\ensuremath{\sigma^{\text{send}}}}
\newcommand \setActionPropose{\ensuremath{\sigma^{\text{propose}}}}

\title{Rationals vs Byzantines in Consensus-based Blockchains}

\author{Yackolley Amoussou-Guenou$^{\ddagger,\star}$, Bruno Biais$^{\dagger}$, \\Maria Potop-Butucaru$^\star$, Sara Tucci-Piergiovanni$^\ddagger$\\~\\
	$^\ddagger$CEA, LIST, PC 174, Gif-sur-Yvette, 91191, France\\
	$^\star$Sorbonne Universit\'e, LIP6, CNRS, UMR 7606, Paris, France \\
	$\dagger$HEC Paris and Toulouse School of Economics, CNRS (TSM-Research), France\\
}

\date{}
\usepackage{placeins}

\begin{document}
	
			%For the pseucode line numbers and labels.
			\newcounter{linecounter}
			\newcommand{\linenumbering}{(\arabic{linecounter})}
			\renewcommand{\line}[1]{\refstepcounter{linecounter}
				\label{#1}
				\linenumbering}
			\newcommand{\resetline}{\setcounter{linecounter}{0}}
\maketitle

\begin{abstract}
	% !TeX root = main.tex
In this paper we analyze from the game theory point of view Byzantine Fault Tolerant  
blockchains when processes exhibit  rational or Byzantine behavior.
Our work is the first to model the Byzantine-consensus based blockchains
 as a committee coordination game.   
 Our \emph{first} contribution is to offer a game-theoretical methodology to analyse equilibrium interactions between Byzantine and rational committee members in  Byzantine Fault Tolerant blockchains. Byzantine processes seek to inflict maximum damage to the system, while rational processes best-respond to maximise their expected net gains. Our \emph{second} contribution is to derive conditions under which consensus properties are satisfied or not in equilibrium. When the majority threshold is lower than the proportion of Byzantine processes, invalid blocks are accepted in equilibrium. When the majority threshold is large, equilibrium can involve coordination failures, in which no block is ever accepted. However, when the cost of accepting invalid blocks is large, there exists an equilibrium in which blocks are accepted iff they are valid.

\end{abstract}
\section{Introduction} \label{sec:intro}
% !TeX root = main.tex

Since the publication of Nakamoto's white paper \cite{bitcoin} proposing the Proof-of-Work protocol, Bitcoin, thousands of blockchains have been created. At the operational level, a blockchain maintains an evolving list of ordered blocks. Each block consists of one or more transactions that have been verified by the system members. POW blockchains, however, consume excessive amounts of energy. This motivated tremendous efforts to propose alternatives protocols. 

Byzantine-consensus based blockchains offer an alternative which has the advantage of being economical and offering strong consistency garanties  \cite{ADLPT18}.  In Byzantine-consensus based blockchains such as HoneyBadger, HotStuff or Tendermint \cite{solidus,Tendermint18,PeerCensus, Bitcoin-NG,BizCoin,honeybadger16,hotstuff18}  a subset of deterministically selected processes, executes an instance of PBFT-consensus to decide on the next block to append.  These protocols strive to satisfy the following properties:   \textit{Termination:} every non-Byzantine process decides on a value (a block); \textit{Agreement:} if there is a non-Byzantine process that decides a value $B$, then all the non-byzantine processes decide $B$;  \textit{Validity\cite{redbelly17}:} a decided value by any non-Byzantine process is valid, it satisfies the predefined predicate.  

While Byzantine consensus  \cite{LSP82} is one of the best understood and formalized building blocks in distributed computing, blockchains systems revive this line of research in several respects: First, traditional Byzantine consensus has been analyzed only in systems where processes were either correct (verify their specification) or Byzantine (arbitrarily deviate from their specification). Blockchain systems bring on the scene a third type of player: rational players who take actions only if these actions increase their profits. Understanding the performance and limits of Byzantine-consensus based blockchains with rational players is the goal of the current work. Our focus on rational players is in line with analyses of blockchain systems conducted by economists. Economists, however, have not considered Byzantine participants yet. Thus, our work endeavours to combine and unify computer science and economics approaches. Second, traditional Byzantine consensus analyses have not studied the choice and consequences of the way in which participants are rewarded. In this work we address the case of rewarding only the participants to the consensus.  

\paragraph{Our contribution.}
Our contribution is twofold. First, we offer a methodology to analyse Byzantine consensus based blockchain protocols as a game between rational and Byzantine players. Two key aspects of the game, for rational players, are the cost of blocks verification and the cost of networking. Block verification is crucial since appending non verified blocks may have long term costs (e.g. double spending, collapse of the system etc.). Networking (participating to the agreement protocol by voting in favor of correct blocks) also has tremendous impact on system welfare: If participants don't vote, this can block the system or lead to agreement on invalid blocks. Second, we derive conditions  (on the majority threshold necessary for block acceptance,  $\nu$, and the proportion of Byzantine processes, $f$)
%(on the threshold $\nu$ and proportion of Byzantine processes, $f$) 
under which rational players reach an equilibrium where the consensus properties are guaranteed. Our findings are as follows. When $f \geq \nu$, invalid blocks are accepted, so that validity is not satisfied. When $f < \nu$, while there exists an equilibrium in which validity and termination are satisfied, there also exists an equilibrium in which blocks are never accepted, so that termination is not satisfied. This points to a tension between validity (which requires that the $\nu$ threshold be large enough) and termination (which can be threatened when the $\nu$ threshold is high.)

%\section{Related Work} \label{sec:relatedwork}
% !TeX root = main.tex

\paragraph{Related work. }Blockchains can be roughly divided in consensus-less \cite{bitcoin}  and Byzantine consensus-based blockchains \cite{BizCoin,PeerCensus,solidus, Bitcoin-NG, Tendermint18}.  Byzantine Consensus-based blockchains have the advantage to guarantee strong consistency by running a Byzantine Fault Tolerant protocol \cite{CL99}. In order to use a BFT protocol in an open setting, recent research has been devoted to either find secure mechanisms to select committees of fixed size over time (e.g.  \cite{Algorand17},\cite{Ouroboros18}) and/or to propose incentives to promote participation \cite{solidus}. Most of the proposals, however, assume participants as either honest or Byzantine, lacking to thoroughly explore the effect of rational participants.  In this line of work, Solidus \cite{solidus} is the first to consider rational processes by proposing an incentive-compatible BFT protocol for blockchains. Solidus introduces interesting incentive mechanisms, however, the paper lacks a game theoretic analysis of them. 

While addressing a slightly different protocol, \cite{MJMF18} is the closest work to ours.
In this protocol multiple committees run in parallel to validate a non-intersecting set of transactions (a shard). A non-cooperative static game approach for the intra-committee protocol is taken leading to the result that rational agents can free-ride when rewards are equally shared. The main aspect of our analysis that is new and different from \cite{MJMF18} is the following: we have a dynamic (not static) multi-round analysis, of a problem in which some participants are Byzantine and some blocks can be invalid (and costly for rational if accepted). In that context, there is a situation in which in equilibrium rational agents are pivotal, because if they do not check the block validity this will create the risk of having an invalid block accepted. It is because they are pivotal that they do not free ride. Moreover, we discuss equilibria in relation to formal consensus properties -- Termination vs Validity --, which represents a novelty.  

In the realm of consensus-less blockchains, as Bitcoin, many works used rational arguments to prove thresholds on the fraction of honest nodes needed to guarantee security properties \cite{ES18,SSZ16}. %, for instance, establishes a two-third honest (computing power) majority to be resilient to the withholding attack (malicious nodes mining a secret chain to replace the main public chain). 
These works establish very pessimistic thresholds while in practice Bitcoin works even if the honest majority assumption does not hold \cite{BGMTZ18}. Following this observation, \cite{BGMTZ18} proposes a rational analysis of Bitcoin based on the rational design protocol framework \cite{GKMTZ13}. The proposed game, with respect to ours, is at an upper level of abstraction, proposing a two-player zero-sum game between the protocol designer and the adversary. Our game models instead the behavior of protocol participants, that can be rational, evolving in an environment with Byzantine processes. Moreover, our work targets consensus-based blockchain unlike \cite{BGMTZ18}.

With only rational players, \cite{BBBC19} models Bitcoin as a coordination game.
%and there exists multiple Nash equilibria, 
Similar to the work in \cite{BBBC19}, our analysis shows that the protocol in consensus-based blockchains is a coordination game. Additionally, we consider Byzantine players, and show that \emph{Termination} can be violated when coordination failures occur.
\cite{Saleh19} uses a game theoretic approach to study consensus-less Proof-of-Stake Blockchains, and shows that the Nothing at Stake problem is mitigated because players with large stakes on the main chain prefer not to add blocks on forking branches, lest it should reduce the strength of the main chain, and thus the value of their stakes. The environment considered in \cite{Saleh19} differs from ours, since the study in \cite{Saleh19} does not consider consensus-based blockchains, nor Byzantine players.

%Blockchains fall in two main classes, forkable and not forkable blockchains. 
%Forkable blockchains rely on two main ingredients, a local cryptographic mechanism, as proof-of-work in Bitcoin or VSS in Alogrand, which designates the generator for the next block,  and a reliable broadcast to diffuse the generated block. Conflicting blocks for the same height are possible, even if all the processes are correct, in a non-synchronous system (to develop  more). 
%The non-forkable class of blockchain use traditional Byzantine consensus algorithms in order to guarantee strong consistency. Tendermint is a variant of PBFT in which the selection mechanim determines a committee that coordinates through a PBFT instance to decide for the next block.

%Incentives are central in both classes to guarantee the security in the system. Unfortunately, most of the security analyses focused on the characterization of nodes as honest and malicious, leaving aside rational behavior. 

\section{Blockchain Consensus with Rational Players}
	\subsection{System Model} \label{sec:model}
		% !TeX root = main.tex

We consider a  system composed of a finite and ordered set $\Pi$, called \emph{committee}, of synchronous sequential processes or players, namely $\Pi = \{p_1, \dots, p_n\}$ where process $p_i$ is said to have index $i$. 
In the following, we refer to process $p_i$ by it index, say process $i$. Hereafter, the words \textquotedblleft player\textquotedblright\ and \textquotedblleft process\textquotedblright\ are taken to have the same meaning. 

\paragraph{Communication.}  We assume that each process evolves in rounds.
A \emph{round} consists of one or more 
%(communication) steps, 
phases,
and each 
%step 
phase
is divided into three sequential 
%phases, 
steps,
in order: the send, the delivery and the compute 
%phase. 
step.
We assume that the send 
%phase 
step
is atomically executed at the beginning of the 
%step 
phase
and the compute 
%phase 
step
is atomically executed at the end of the 
%step. 
phase.
The 
%step 
phase
has a fixed duration that allows collecting all the messages sent by the processes at the beginning of the 
%step
phase 
during the delivery 
%phase.
step.
% : (i) a \emph{Send} phase, where the process broadcasts messages computed during the last round, or a default messages for the first round; (ii) a \emph{Delivery} phase where the process collect messages sent during the current and previous rounds; and (iii) a \emph{Compute} phase where the process uses the messages delivered to change its state.
At the end of a 
%step
phase 
a process exit from the current 
%step
phase 
and starts the next one.
%Each communication step last a certain duration, we consider the Send and the Compute phase as being atomic, they are executed instantaneously, but not the Delivery phase.
%In a synchronous network, we assume the the duration of the Delivery phase, and so of the round is $\delta$.
The processes communicate by sending and receiving messages through a broadcast primitive. 
%A process $p_i$ sends a message by invoking $\textsf{broadcast} (\langle TAG,m \rangle)$, where $TAG$ is the type of the message, and $m$ its content. To simplify the presentation, it is assumed that a process can send messages to itself. 
%
%A process $p_i$ receives a message by executing the primitive $\textsf{delivery}()$.
Messages are created with a digital signature, and we assume that digital signatures cannot be forged. When a process $i$ delivers a message, it knows the process $j$ that created the message. We assume that messages cannot be lost. 

\paragraph{Processes Behavior.}
In this paper we consider a variant of the BAR model \cite{AACDMP05} where processes are either  \emph{rational} or \emph{Byzantine}.  \emph{Rational processes} are self-interested and seek to maximize their expected utility. They will deviate from a prescribed (suggested) protocol if and only if doing so increases their expected utility. Their objective function must account for their costs (e.g., sending messages) and benefits (e.g., reward of a block) for participating in a system. In line with \cite{AACDMP05}, the objective of Byzantine processes is prevent the protocol from achieving its goal, and to reduce the rational processes utility, no matter the cost. We denote by $f$ the number of Byzantine processes in the network. We assume symmetric Byzantines, their behaviour is perceived identically by all non Byzantine processes. That is, a message sent by a Byzantine process and received by a non-Byzantine process in a given phase is received by all non-Byzantine processes in the same phase.

	\subsection{Byzantine Consensus based Blockchain}\label{sec:problemDef}
%	\subsection{Consensus Specification}\label{sec:problemDef}
		% !TeX root = main.tex

%In  consensus-based blockchains \cite{solidus,Tendermint18,PeerCensus, Bitcoin-NG,BizCoin,honeybadger16,hotstuff18} ,  a subset of deterministically selected processes,   executes an instance of  PBFT-consensus to decide on the next block to append.
%All these protocols strive to satisfy 
Consensus-based blockchains should satisfy the following consensus properties: 
\begin{itemize} 
	\item\textbf{\textit{Termination:}} every non-Byzantine process decides on a value (a block); 
	\item\textbf{\textit{Agreement:}} if there is a non-Byzantine process that decides a value $B$, then all the non-byzantine processes decide $B$;  
	\item\textbf{\textit{Validity\cite{redbelly17}:}} a decided value by any non-Byzantine process is valid, it satisfies the predefined predicate.
\end{itemize}
%In this work, we study the whether or not \emph{termination} and \emp{validity} can be guaranteed with the presence of Rational processes.

Let us note that the above properties must hold also for systems prone to  rational behavior.

To implement the above specification in Consensus-based blockchains, for each height $h>0$ of the blockchain, a Consensus instance is run
inside a committee selected for the given height.  In this paper we analyze a very general protocol,
inspired by \cite{solidus,Tendermint18,PeerCensus, Bitcoin-NG,BizCoin,honeybadger16,hotstuff18}, a variant of PBFT.
In this protocol, a proposer proposes a value, i.e. a block, and the other members of the
committee will check the validity of the value. If the value is valid, then
they will vote for it and will announce their vote through a message to the other members. Votes
are collected and if a given threshold is reached, then the value is decided, otherwise a new
proposer will propose another block and the procedure restarts.
%	\subsection{Principle of the Protocols}
		% !TeX root = main.tex

In this work, we study a protocol	(for rational players) which in some equilibria implements the consensus. For the sake of clarity, we first present a prescribed protocol, and then the actions of the rational processes. % can have.
%A rational player is a process that executes a prescribed protocol only if the protocol matches her own interest. 
If a rational player does not deviate from a given prescribed protocol, we can consider it as a correct process. % -- a process that does not deviate from its specification. 

\paragraph{The prescribed protocol.} The protocol proceeds in rounds. For sake of simplicity we consider the height $k$ of the blockchain passed as parameter to the protocol. 
Algorithm \ref{fig:protocolAltruistic} presents the pseudo-code of the protocol.
%, and is similar to the protocol presented in \cite{dissecting18}.

% !TeX root = main.tex

\begin{algorithm}[t!] \def\baselinestretch{1} \scriptsize\raggedright
	 \begin{algorithmic}[1] \SHORTSPACE 
		\INIT{} 
			\STATE $vote := \nil$ 
			\STATE $\currRound := 1$ \COMMENT{Current round number}
		 	\STATE $decidedValue :=\nil $  
	 	\ENDINIT 
	 	
		\SPACE \ROUND{PROPOSE($\currRound$)}\label{Pprop-beginPcons}
			\SENDP{}
				\IF{$i == \isProposer(\currRound, \currHeight)$}
				    \STATE $proposal\assign \createValidValue(\currHeight)$\COMMENT{The proposer of the round generates a block, i.e. the value to be proposed}
					\STATE \Broadcast\ $\li{\Proposal, \currHeight, \currRound, proposal}$
				\ENDIF
			\ENDSENDP
			\DEL{}
				\STATE \textbf{delivery} $\li{\Proposal,\currHeight, \currRound, v}$ from $\coord(\currRound)$ \COMMENT{The process collects the proposal}
			\ENDDEL
			\COMP{}
				\IF{$\isValid(v)$}
					\STATE $vote \assign v$  \COMMENT{If the delivered proposal is valid, then the process sets a vote for it}
				\ENDIF
			\ENDCOMP
		\ENDROUND
		
		\SPACE \ROUND{VOTE($\currRound$)}\label{Vote-beginPcons}
			\SENDP{}
				\IF{$\vote \neq  nil$}
					\STATE\Broadcast \ $\li{\Precommit_i,\currHeight,\currRound,vote}$ \COMMENT{If the proposal is valid, the process sends the vote for it to all the validators}
				\ENDIF	
			\ENDSENDP
			
			\DEL{}
				\STATE \textbf{delivery} $\li{\Precommit,\currHeight, \currRound, v}$ \COMMENT{The process collects all the votes for the current height and round}
			\ENDDEL
			
			\COMP{}
				\IF{$|\li{\Precommit, \currHeight, \currRound, v}| \ge \nu \land decidedValue = \nil \land  vote \neq nil \land  vote = v$}  
					\STATE $decidedValue  \assign v; $  $\textbf{exit}$ \COMMENT{The valid value is decided if the threshold is reached}
				 \ELSE
				 	\STATE $vote  \assign \nil$
				 	\STATE $\currRound \assign \currRound+1$	
				\ENDIF
			\ENDCOMP
		\ENDROUND
	\end{algorithmic} 
	\caption{Prescribed Protocol for a given height $\currHeight$ at any process $i$}
	\label{fig:protocolAltruistic} 
\end{algorithm}

%\paragraph{Description of Algorithm \ref{fig:protocolAltruistic}}
%	\YAG{Description a faire}

%We also consider the broadcast function to disseminate the messages to the whole blockchain network, so to include committee members as well. The delivery of the protocol's messages will be activated only at those processes being part of the current committee.
For each round $t$ a committee member is designated as the proposer for the round in a round robin fashion. The \texttt{isProposer}($t,k$) function returns true only if the process is the proposer for the current round  (line 7). The function, by taking as parameter the current height, only returns  true if the proposer is part of the current committee, deterministically selected on the basis on the information contained in the blockchain up to $k$ (the actual selection mechanism is out of the scope of the paper).  
Each round is further divided in two phases: the PROPOSE and the VOTE phase. %Moreover, each phase is dived in three steps: the send, the delivery and the compute step. %We assume that the send step is atomically executed at the beginning of the phase and the compute step is atomically executed at the end of the phase. The phase has a fixed duration that allows collecting all the messages sent by the processes at the beginning of the phase during the delivery step. %Sent messages cannot be lost and are delivered by the end of the phase, but Byzantine and rational processes might deviate from the protocol not sending a prescribed message (more details later in the corresponding pseudo-codes).

During the PROPOSE  phase,  the proposer of the round uses the function \texttt{createValidValue}$(k)$ to generate a block. Because a valid block must include the identifier of the $k^{th}$ block in the blockchain, the height $k$ is passed as parameter (line 8). Once the block is created, a message broadcasting the proposal is sent (line 9). 
At line 10 the proposal is received through a delivery function. Each process checks if the proposal is a valid value (line 13). If so,  the process sets its vote to the value (line 14). 
%Let us note that we assume that the proposal is always sent, the only bad thing that can happen is that the sent proposal is not valid (more later in the rational and Byzantine pseudo-codes).\ST{Note that this is not a restrictive assumption, since if the proposal is not sent and then not received at the end of the phase, the vote remains to nil and all the non-Byzantine processes would move the next round, but we did not really look at that for the rational behavior.}

During the VOTE phase, any process that set its vote to the current valid proposal sends a message (of type vote) to the other members of the committee (line 18). During the delivery step, sent messages are collected by any process. During the compute step each process verifies if a quorum of $\nu$ votes for the current proposal has been reached. Let us note that $\nu$, the majority threshold is a parameter here, because it is the object of our study to establish the quorum $\nu$  in presence of rational and Byzantine processes. If the quorum is reached, if the process voted for the value and did not already decided for the current height, then it decides for the current proposal (line 23) and the protocol ends. If the quorum is not reached, then a new round starts (line 26).   
%\input{algoAltruistic}

%Let us note that the protocol in an environment assuming only correct and Byzantine processes with symmetric communication trivially implements Consensus if $f$, the number of Byzantine processes, is such that $f<\nu$. If $f\geq \nu$, on the other hand, the Termination property is not guaranteed. The scenario for that is that Byzantine validators might vote for a different value with respect to the one voted by correct processes or a nil value. In that case the correct process will not decide (line 22) and will move in the next round. The scenario can repeat forever. 

Let us note that the protocol in an environment assuming only correct (altruistic) and symmetric Byzantine processes  trivially implements consensus if $f$, the number of Byzantine processes, is such that $f<\nu$. If $f\geq \nu$, on the other hand, the Termination property is not guaranteed. The scenario for that is that Byzantine validators might vote for a different value with respect to the one voted by correct processes or a nil value. In that case the correct process will not decide (line 22) and will move in the next round. The scenario can repeat forever. 
% !TeX root = main.tex

\begin{figure}
	\center
%	\resizebox{\linewidth}{!}{
	\begin{tikzpicture}[grow=right, sloped]
	\node[bag]{}
%	child{
%	\node[bag]{Proposer is rational}
%	child{
%	node[bag] {Proposal valid}%{Bag 1 $4W, 3B$}
	    child {
	        node[bag] {Does not check the validity (line 21)}        
	            child {
	                node[bag] {No Validity Information (line 6)}
	                child{
	                	node[end, label=right: Not Send (line 24)] {}
	            		edge from parent 
	                } child{
	                	node[end, label=right: Send (lines 24 -- 30)] {}
	          		edge from parent 
	                }
	                edge from parent
	            }
	            edge from parent 
	%            node[above] {$W$}
	%            node[below]  {$\frac{4}{7}$}
	    }
	    child {
	        node[bag] {Checks the validity (lines 21--23)}        
	            child {
	                node[bag] {Not Valid}
	                child{
	                	node[end, label=right: Not Send (line 24)] {}
	            		edge from parent 
	                } child{
	                	node[end, label=right: Send  (lines 24 -- 30)] {}
	          		edge from parent 
	                }
	                edge from parent
	            }        
	            child {
	                node[bag] {Valid}
	                child{
	                	node[end, label=right: Not Send (line 24)] {}
	            		edge from parent 
	                } child{
	                	node[end, label=right: Send  (lines 24 -- 30)] {}
	          		edge from parent 
	                }
	                edge from parent
	            }
	        edge from parent
	    }
%	    edge from parent
%	    }
%	    edge from parent
%	    }
%	    
%	    child{
%	    node[bag]{Proposer is symmetric Byzantine}
%	    child{
%	    node[bag] {Proposal inalid}%{Bag 1 $4W, 3B$}
%	    child {
%	        node[bag] {Does not check the validity}        
%	            child {
%	                node[bag] {No Validity Information}
%	                child{
%	                	node[end, label=right: Not Send] {}
%	            		edge from parent 
%	                } child{
%	                	node[end, label=right: Send] {}
%	          		edge from parent 
%	                }
%	                edge from parent
%	            }
%	            edge from parent 
%	%            node[above] {$W$}
%	%            node[below]  {$\frac{4}{7}$}
%	    }
%	    child {
%	        node[bag] {Checks the validity}        
%	            child {
%	                node[bag] {Not Valid}
%	                child{
%	                	node[end, label=right: Not Send] {}
%	            		edge from parent 
%	                } child{
%	                	node[end, label=right: Send] {}
%	          		edge from parent 
%	                }
%	                edge from parent
%	            }        
%	            child {
%	                node[bag] {Valid}
%	                child{
%	                	node[end, label=right: Not Send] {}
%	            		edge from parent 
%	                } child{
%	                	node[end, label=right: Send] {}
%	          		edge from parent 
%	                }
%	                edge from parent
%	            }
%	        edge from parent
%	    }
%	    edge from parent
%	    }
%	    edge from parent
%	    } 
	    ;
	\end{tikzpicture}
%	}
	\caption{Decision tree of process $i$}
	\label{tree:VS}
\end{figure}
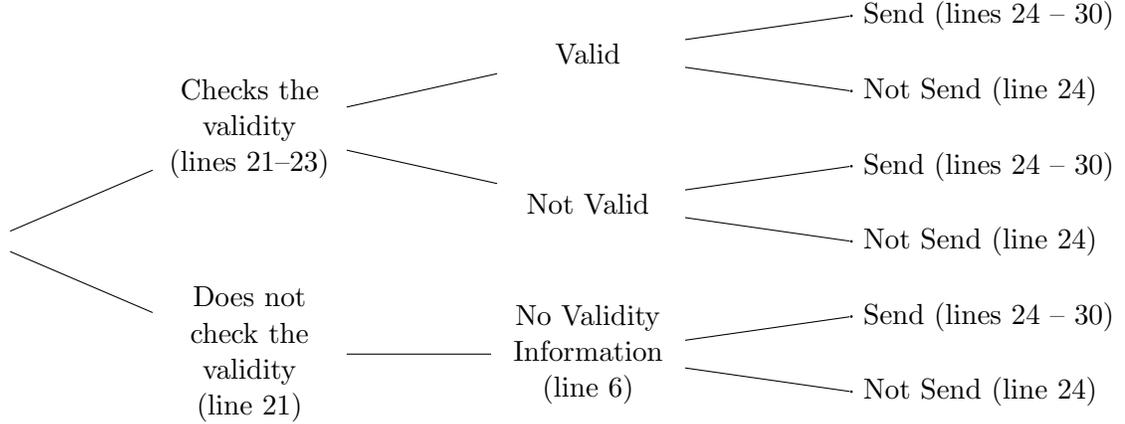

In the following we detail the pseudo-code for a rational processes  shown in Algorithm \ref{fig:protocolVS}.  The rational process will try to maximize its payoff by choosing to undertake or not the actions defined in its action space. We consider the choice of : (i) proposing or not a valid block, (ii) checking or not the validity of a block and (iii) sending or not the vote for a proposed block. The decision tree for the process $i$ is shown in Figure \ref{tree:VS}.

Let us consider now rational processes. The rational process will try to maximize its payoff by choosing to undertake or not some actions, defined in Section \ref{sec:gameModel}.
Intuitively, we consider the choice of : (i) proposing or not a valid block, (ii) checking or not the validity of a block and (iii) sending or not the vote for a proposed block. We consider that the actions of checking the validity of the block and the action of sending the message (of type vote) have a cost.%, respectively $c_{check}$ and $c_{send}$. 

\paragraph{Protocol of the rational processes.}
Rational processes choices are explicitly represented in the pseudo-code (Algorithm \ref{fig:protocolVS}) by dedicated variables, namely, $action^{propose}$, $action^{check}$, and $action^{send}$, defined at lines $5-7$. Each action, initialized to $nil$, can take values from the set $\{0,1\}$. Those values are set by calling the functions $\setActionPropose_i$, $\setActionCheck_i$, and $\setActionSend_i$, respectively, returning the strategy for the process $i$. 

Strategy $\setActionPropose_i$ determines if the proposer $i$ chooses to produce a valid proposal or an invalid one (lines 12-16). In both cases the proposal is sent in broadcast (line 17). 

Strategy $\setActionCheck_i$ determines if the receiving process chooses to check the validity of the proposal or not, which is a costly action. If the process chooses to check the validity (line 22), it will also update the knowledge it has about the validity of the proposal and it will pay a cost $c_{check}$. If otherwise, the process keeps not knowing if the proposal is valid or not  ($validValue[\currRound]$ remains set to $\unknown$). Note that this value remains set to $\unknown$ even if the process is the proposer. This is because we assumed, without loss of generality, that checking validity has a cost and that the only way of checking validity is by executing the $\isValid(v)$ function.

% !TeX root = main.tex

\begin{algorithm}[t!] \def\baselinestretch{1} \scriptsize\raggedright
	 \begin{algorithmic}[1] %\SHORTSPACE 
		\INIT{} 
			\STATE $vote := \nil$ 
			\STATE $\currRound := 1$ \COMMENT{Current round number}
			\STATE $decidedValue :=\nil $  
             \STATE $\actionPropose:= \nil$			
			\STATE $\actionCheck := \nil$
			\STATE $\actionSend:= \nil$
			\STATE $validValue [] :=\{ \unknown,\unknown, \ldots,  \unknown  \}$ \COMMENT{$validValue[\currRound] \in \{\unknown, 0, 1\}$}
			\SHORTSPACE 
			
		\ENDINIT

		\SPACE \ROUND{PROPOSE($\currRound$)}\label{Pprop-beginPcons}
			\SENDP{}
				\IF{$i == \isProposer(\currHeight,\currRound)$}
				\STATE $\actionPropose \assign \setActionPropose_i()$ \COMMENT{$\setActionPropose_i \in \{\false,\true\}$ sets the action of proposing a valid block or an invalid one}
			\IF{$\actionPropose == \true$}
	 			\STATE $proposal \assign \createValidValue(\currHeight)$
	 		\ELSIF{$\actionPropose == \false$}
	 			\STATE $proposal \assign \createInvalidValue()$
	 		\ENDIF
					\STATE \Broadcast\ $\li{\Proposal,\currHeight, \currRound, proposal}$ 
				\ENDIF
			\ENDSENDP
			\DEL{}
				\STATE \textbf{delivery} $\li{\Proposal,\currHeight, \currRound, v}$ from $\coord(\currHeight, \currRound)$
			\ENDDEL
			\COMP{}
				\STATE $\actionCheck \assign \setActionCheck_i()$ \COMMENT{$\setActionCheck_i \in \{0,1\}$ sets the action of checking or not the validity of the proposal}
				\IF{$\actionCheck== \true$}
					\STATE $\validValue[\currRound] \assign \isValid(v)$ \COMMENT{The execution of  $\isValid(v)$ has a cost $c_{check}$}
				\ENDIF
				\STATE $\actionSend \assign \setActionSend_i(\validValue)$ \COMMENT{$\setActionSend_i: \{\unknown, 0, 1\} \to \{0,1\}$ sets the action of sending the vote or not}
				\IF{$\actionSend == \true$}
					\STATE $vote \assign v$ \COMMENT{The process decides to send the vote, the proposal might be invalid}
				\ENDIF
			\ENDCOMP
		\ENDROUND
		
		\SPACE \ROUND{VOTE($\currRound$)}\label{Vote-beginPcons}
			\SENDP{}
				\IF{$\vote \neq  nil$}
					\STATE\Broadcast \ $\li{\Precommit_i,\currHeight,\currRound,vote}$ \COMMENT{The execution of  the broadcast has a cost $c_{send}$}
				\ENDIF	
			\ENDSENDP
			
			\DEL{}
				\STATE \textbf{delivery} $\li{\Precommit,\currHeight, \currRound, v}$ \COMMENT{The process collects all the votes for the current height and round}
			\ENDDEL
			
			\COMP{}
				\IF{$|\li{\Precommit, \currHeight, \currRound, v}| \ge \nu \land decidedValue = \nil  \land  vote \neq nil \land  vote = v $}
					\STATE $decidedValue = v; $  $\textbf{exit}$
				 \ELSE
				 	\STATE $vote  \assign \nil$
				 	\STATE $\currRound \assign \currRound+1$	
				\ENDIF
			\ENDCOMP
		\ENDROUND
	\end{algorithmic} 
	\caption{Pseudo-code for a given height $\currHeight$ modelling the rational process $i$'s behavior }
	\label{fig:protocolVS} 
\end{algorithm}

%\paragraph{Description of Algorithm \ref{fig:protocolVS}}
%	\YAG{Description a faire}

Note that, as defined in Section \ref{sec:gameModel}, strategy $\setActionSend_i$ depends on the knowledge the process has about the validity of the proposal. The strategy determines if the process chooses to send its vote for the proposal or not (line 24-30). If the processes chooses to send a message for the proposal it will pay a cost $c_{send}$. 

Let us note that the rational player that did not check the validity of the block could decide an invalid value if more than $\nu$ other processes have done the same and the proposed block is invalid. 
%Intuition for that is that the rational player would guess that others will check the validity, so that it will free-ride without paying the cost. 
%Another scenario that can happen in case of a valid proposed block is that only less than $\nu$  processes will send the vote, in that case Termination is not guaranteed. 
%Exact conditions about these strategies are determined in the sequel. 
%Another scenario (path 3 in Figure \ref{tree:VS}) could lead to decide an invalid value, even though the rational player checked the validity and she knows that the block is invalid. This path, however, is never profitable for the rational player and will never be executed, as shown in the  rest of the paper. 

%\newpage
%More in details : $\texttt{isValid}(v)$ (line 23), i.e. checking the validity of the proposal, and $\Broadcast \ \li{\Precommit_i,h,\currRound,vote}$ (line 30), sending a vote for the current proposal. 

We now define the game that represent the protocol.

	\subsection{Byzantine-Rational Game}\label{sec:gameModel}
		% !TeX root = main.tex

Recall that out of the $n
$ players, $f\geq1$ are Byzantine, while $n-f$ are rational. Each player $i$
privately observes its own type, $\theta_{i}$, which can be Byzantine
($\theta_{i}=\theta^{b}$) or rational ($\theta_{i}=\theta^{r}$).\footnote{If
player's type was observable (i.e., if Byzantine processes were detectable in
advance) there would be a trivial solution to preclude them from harming the
system: forbidding their participation.}

\paragraph{Action space.}
%If round $t\in\{1,...,n\}$ is reached, the proposer of round $t$ proposes a block. 
As proposer, the player
decides whether to propose a valid block or to propose an invalid block.

Then, at each round $t$, each player first decides whether to check the
block's validity or not (at cost $c_{check}$), and second decides whether to
send a message (at cost $c_{send}$) or not.

\paragraph{Information sets.}
At the beginning of each round $t>1$, the information set of the player,
$h_{i}^{t}$, includes the observation of the round number $t$, the player's
own type $\theta_{i}$, as well as the observation of what happened in previous
rounds, namely (i) when the player decided to check validity, the knowledge of
whether the block was valid or not, (ii) how many messages were sent, and (iii)
whether a block was accepted or not. At round 1, $h_{i}^{1}$ only includes the
player's private information about its own type, $\theta_{i}$.

Then, in each round $t>1$, the player decides whether to check the validity of
the current block. At this point, denoting by $b_{t}$ the block proposed at
round $t$, when the player does not decide to check validity $\isValid(b_{t})$ 
is the null information set, while if the player decides to check,
$\isValid(b_{t})$ is equal to 1 if the block is valid and 0 otherwise.
So, at this stage the player's information set becomes%
\[
H_{i}^{t}=h_{i}^{t}%\text{ }%
%TCIMACRO{\tbigcup }%
%BeginExpansion
%{\textstyle\cup}
\cup
%EndExpansion
\isValid(b_{t}),\text{ }%
\]
which is $h_{i}^{t}$ augmented with the validity information 
player $i$ has about $b_t$, the proposed block.

\paragraph{Strategies.}

At each round $t\geq1$, the strategy of player $i$ is a mapping from its
information set into its actions. If the agent is selected to propose the
block, its choice is given by 
%$\sigma_{i}^{propose}(h_{i}^{t})$. 
$\setActionPropose_{i}(h_{i}^{t})$. 
Then, at the
point at which the agent can decide to check block's validity, its strategy is
given by 
%$\sigma_{i}^{check}(h_{i}^{t})$. 
$\setActionCheck_i(h_{i}^{t})$. 
Finally, after making that decision,
the player must decide whether to send a message or not, and that decision is
given by %$\sigma_{i}^{send}(H_{i}^{t})$.
$\setActionSend_i(H_{i}^{t})$. 

\paragraph{Reward and cost from adding blocks.}

In this paper we study the case in which, when a block is accepted, only the
processes which sent a message are rewarded (and receive $R$). 
%In the next
%version of the paper we will study equilibria arising when all participants
%are rewarded.
 In addition, we assume that when an invalid block is accepted,
all rational players incur cost $\kappa$.

%\subparagraph{Assumption 1:}

\textit{In this work we make the following assumption. The reward, $R$, to the players when a block is accepted is larger than the
cost of checking validity, $c_{check}$, which in turn is larger than cost of sending, $c_{send}$, a
message. But the reward obtained when a block is accepted is smaller than the
cost of accepting an invalid block, $\kappa$.  That is, $\kappa>R>c_{check}>c_{send}$.}

\paragraph{Objective of rational players.}
Let $T$ be the endogenous round at which the game stops. If a block is
accepted at round $t\leq n$, then $T=t$. Otherwise, if no block is accepted,
$T=n$. In the latter case, the \textit{termination} property is not satisfied.

At the beginning of the first round, the expected gain of rational player $i$ is:%

\[
U_{i}=E\left[
\begin{array}
[c]{c}%
(R\ast\ind_{(\setActionSend_i(H_{i}^{T})=1)}\ast\ind%
_{(\text{block accepted at }T)}-\kappa\ind_{(\text{invalid block
accepted})})\\
-\sum_{t=1}^{T}\left(  c_{check}\ind_{\setActionCheck_i(h_{i}%
^{t})=1)}+c_{send}\ind_{(\setActionSend_i(H_{i}^{t})=1)}\right)
\end{array}
|h_{i}^{1}\right]  ,
\]
where $\ind_{(.)}$ denotes the indicator function, taking the value 1
if its argument is true, and 0 if it is false.

Then, at the beginning of round $t>1$, if $T\geq t$, the continuation payoff
of the rational player with information set $h_{i}^{t}$ is%
\[
W_{i,t}(h_{i}^{t})=E\left[
\begin{array}
[c]{c}%
(R\ast\ind_{(\setActionSend_i(H_{i}^{T})=1)}\ast\ind%
_{(\text{block accepted at }T)}-\kappa\ind_{(\text{invalid block
accepted})})\\
-\sum_{s=t}^{T}\left(  c_{check}\ind_{(\setActionCheck_i(h_{i}%
^{s})=1)}+c_{send}\ind_{(\setActionSend_i(H_{i}^{s})=1)}\right)
\end{array}
|h_{i}^{t}\right]  ,
\]

\paragraph{Objective of Byzantine players.}

%The objective of Byzantine players is to prevent the protocol from achieving
%its goal, and to inflict maximum damage.

% To capture this notion, we will
%assume, in the next draft that the Byzantine players choose their strategies
%to minimise the gains of the rational players. So, at round $t=1$, each
%Byzantine player will choose its strategy to maximise $-U_{i},$while at time
%$t>1$ it will maximise $-W_{i,t}$. In the present, preliminary, draft, we
%don't derive endogenously the strategies of the Byzantine players, but assume
%they are as follows:
%\subparagraph{Assumption 2:}
In the current paper we assume the following:
\textit{Byzantine processes 1) as proposers, propose invalid blocks, and 2)
when receiving a proposed block, check the blocks' validity and send a message
if and only if the block is invalid.}

We conjecture, the above strategies  will turn out to be the
optimal strategies of the Byzantine players, minimizing $W_{i,t}$ in equilibrium.

\paragraph{Equilibrium concept.}

Since we consider a dynamic game, with asymmetric information, the relevant
equilibrium concept is Perfect Bayesian Equilibrium \cite{FT91}, intuitively defined as follows:

%\subparagraph{Definition:}

\textit{A Perfect Bayesian equilibrium is such that all players 1) choose
actions maximizing their objective function, 2) rationally anticipate the
strategies of the others, and 3) draw rational inferences from what they
observe, using their expectations about the strategies of the others and Bayes
law, whenever it applies.}

A Perfect Bayesian Equilibrium (PBE) is a Nash equilibrium \cite{Nash51}, so players
best-respond to one another. It imposes additional restrictions, to take into
account the fact that the game is dynamic and that players can have private
information, and therefore must draw rational inferences, from their
observation of actions and outcomes. Rationality of inferences in PBE implies
that (i) each player has rational expectations about the strategies of the
others, and (ii) each player's beliefs are consistent with Bayes law, when
computing probabilities conditional on events that have strictly positive
probability on the equilibrium path. Perfection in PBE implies that, at each
node starting a subgame the players' strategies form a Nash equilibrium of
that subgame. In this context, to show that a strategy is optimal it is
sufficient to show that it dominates any one-shot deviation \cite{Blackwell65}.

\paragraph{Problem Definition}
In this work, we explore the behavior of rational players that could not validate the block -- because checking validity has a cost -- and conditions (the majority threshold $\nu$ and proportion of Byzantine processes) under which rational players reach an equilibrium where consensus properties (defined in Section \ref{sec:problemDef}) 
are guaranteed. To do so, in Section \ref{sec:results}, we %next describe the game (Section \ref{sec:gameModel}),
study the equilibria that arise under different conditions.

%Due to space limitations, proofs not in the text are provided in the Appendix.	
	%\section{Problem Definition and Protocols}\label{sec:Problem}
	%	\subsection{Consensus-Based Blockchain in the BAR model}\label{ssec:RCBlockchain}

%\input{protocols}
	
\section{Equilibria for Rational Players}\label{sec:results}
	\subsection{Equilibrium when $f \geq \nu$
	%When the number of Byzantine processes is larger than or equal to the majority threshold
	}\label{ssec:badSetting}
		% !TeX root = main.tex

When the number of Byzantine players is larger than the majority rule, i.e.,
$f\ge\nu$, the validity property is not satisfied, since, when the first
proposer is Byzantine, it proposes an invalid block, and that block is
accepted, as all Byzantine players send messages in its favor. Against that
backdrop, we characterize the strategies of the rational players and state the
equilibrium outcome when $f\ge\nu$.

%\subparagraph{Proposition 1:}
\begin{proposition}
\label{prop1}
\textit{If }$n-f\geq\nu+1$ and $f\geq\nu$\textit{, there exists a Perfect Bayesian
equilibrium in which the strategy of a rational player at any round is the
following:}

\begin{itemize}
\item \textit{As proposer, a rational player proposes a valid block.}

\item \textit{When receiving a proposed block, the rational players do not
check the block validity but send a message.}
\end{itemize}
\end{proposition}
%\bigskip

%\subparagraph{Intuition for Proposition 1:}
%The intuition of Proposition \ref{prop1} is as follows.
The first condition ($n-f\geq\nu+1$) implies that, when all rational players
but one send a message, they meet the majority threshold $\nu$, so the block
is accepted. The second condition ($f\geq\nu$) implies that, when all
Byzantine processes send a message, the block is accepted. Under these
conditions, each rational player understands it is not pivotal: If the block
is invalid, Byzantine players will send messages, so that the block will be
accepted irrespective of the rational player's own action. Moreover, if the
block is valid, Byzantine players will not send messages, but all the other
rational players will, so that the block will be accepted irrespective of the
rational player's action.

Thus rational players understand that they are not pivotal, and that whatever
they do, given the equilibrium behavior of the other rational agents and of
the Byzantine processes; all blocks will be accepted. Consequently, they have
no interest in checking the validity of the block. The only relevant
comparison for them is between their expected gain when they send a message%
\[
R-c_{send}-\frac{f}{n}\kappa
\]
and their expected gain when they do not send a message $-\frac{f}{n}\kappa$.
Since, by assumption, $R>$ $c_{send}$, rational players find it optimal to
send a message. Finally note that, in the equilibrium of Proposition \ref{prop1}, a
block is decided at round 1, so the \textit{termination} property is
satisfied, but, when the proposer is Byzantine, an invalid block is accepted,
so the \textit{validity} property is not satisfied.
		% !TeX root = main.tex

\begin{proofP}
If a rational player is selected to be the proposer, he prefers to propose a
valid block than to propose an invalid block. Indeed, if he proposes an
invalid block, that block will be accepted (since the $f\ge\nu$ Byzantine
players, on checking it and discovering it is invalid, will send a message).
In that case the gain of the proposer is $R-c_{check}-c_{send}-\kappa$. If
instead the rational player proposes a valid block, this block will be
accepted and his gain will be $R-c_{check}-c_{send}$. Now, turn to the actions
of rational players who are not proposers. The equilibrium gain of these
players is%
\[
-c_{send}+R-\frac{f}{n}\kappa.
\]

If instead of playing the equilibrium strategy, a rational player does not
send a message, its expected gain is $-\frac{f}{n}\kappa$, which by 
assumption ($R > c_{send}$) is lower than the equilibrium expected gain.

Another deviation is to check the block's validity and send a message only if
the block is valid, which brings expected gain equal to%
\[
-c_{check}+(1-\frac{f}{n})(R-c_{send})-\frac{f}{n}\kappa.
\]
This is lower than the equilibrium expected gain if%
\[
-c_{send}+R-\frac{f}{n}\kappa>-c_{check}+(1-\frac{f}{n})(R-c_{send})-\frac
{f}{n}\kappa,
\]
which holds since it is equivalent to%
\[
0>-c_{check}-\frac{f}{n}(R-c_{send}).
\]

The other possible deviations are trivially dominated: Checking the block's
validity and sending a message only when the block is invalid, yields expected
gain%
\[
-c_{check}+\frac{f}{n}(R-c_{send}-\kappa),
\]
which is lower than the equilibrium expected gain. Checking the block's
validity and sending a message only when the block is valid yields expected
gain%
\[
-c_{check}+\left(  1-\frac{f}{n}\right)  (R-c_{send})-\frac{f}{n}\kappa,
\]
again lower than the equilibrium expected gain. Checks the validity of the
block and always sending a message yields%
\[
R-c_{send}-\frac{f}{n}\kappa-c_{check},
\]
which is again dominated, as is also checking and not sending, which yields
$-c_{check}-\frac{f}{n}\kappa$.

%QED
\renewcommand{\toto}{prop1} 
\end{proofP}
	
	\subsection{Equilibria when $f < \nu$
	%When the number of Byzantine processes is strictly smaller than the majority threshold
	}\label{ssec:goodSetting}
		% !TeX root = main.tex

%\subparagraph{Proposition 2:}
\begin{proposition}
\label{prop2}
\textit{When }$f<\nu$\textit{\ and }$n-f\geq\nu$\textit{, there exists a 
Nash equilibrium in which rational players never check blocks' 
validity nor send messages, so that no block is ever accepted.}
\end{proposition}
%\bigskip

%\subparagraph{Intuition for Proposition 2:}
%The intuition of Lemma \ref{prop2} is as follows.
Condition $f<\nu$, in Proposition \ref{prop2} implies that Byzantine players cannot
reach the majority threshold on their own. This precludes accepting invalid
blocks. So the \textit{validity} property is satisfied. Unfortunately, the
condition also implies there exists an equilibrium in which the
\textit{termination} property also fails to hold. The intuition is the following:

In Proposition \ref{prop2}, each rational player anticipates that no other player will
send a message when the block is valid.\footnote{Byzantine players send
messages but only when the block is invalid.} In this context, each rational
player knows that, if it were to send a message in favor of a valid block, it
would be the only one to do so. Because the majority threshold $\nu$ is
strictly larger than 1, the block would not be accepted. Therefore sending a
message is a dominated action for the rational player. Thus, the equilibrium
in Proposition \ref{prop2} reflects that rational players' actions are strategic
complements and they must coordinate on sending messages in order to have
valid blocks accepted. Proposition \ref{prop2} shows that, in equilibrium, there can be
a coordination failure, such that no block is ever accepted.\footnote{If
$f=0$, then, with $\nu=1$, there exists a unique equilibrium, in which all
processes check validity and send a message iff the block is valid. In that
equilibrium validity and termination are satisfied. But this obtains only if
there are no Byzantine processes. As soon as $f\geq1$, if $\nu=1$, 
Proposition \ref{prop1} applies and validity is not satisfied.}

		% !TeX root = main.tex

\begin{proofP}
Consider a rational player who anticipates that other rational players will
not send any message at any round. If it follow the equilibrium strategy and
does not send an message, its gain is 0. This must be compared to the gain of
the player if he deviates:

\begin{itemize}
\item If it sends a message without checking its expected gain is
\[
-c_{send}+\Pr(\text{invalid})\ind_{(f=\nu-1)}(R-\kappa).
\]

\item If it checks the block's validity and sends a message only when the
block is valid, its expected gain is%
\[
-c_{check}-\Pr(\text{valid})c_{send}.
\]

\item If it checks the block's validity and sends a message only when the
block is invalid, its expected gain is%
\[
-c_{check}+\Pr(\text{invalid})(\ind_{(f=\nu-1)}(R-\kappa)-c_{send}).
\]

\item If it checks the validity of the block and always sends a message, its
expected gain is%
\[
-c_{send}-c_{check}+\Pr(\text{invalid})\ind_{(f=\nu-1)}(R-\kappa).
\]

\item If it checks and does not send a message, its gain is $-c_{check}$.
\end{itemize}

Clearly, the player is better off following the equilibrium strategy.

%QED
\renewcommand{\toto}{prop2} 
\end{proofP}

Note that the conditions of Proposition \ref{prop2} imply that $f<\frac{n}{2}$, i.e,
there is a strict majority of rational players. Yet, the proposition shows
that such majority is not enough to ensure both termination and validity.

%\bigskip

While there exists an equilibrium in which termination does not obtain, this
does not necessarily imply there is no equilibrium with termination and
validity. To have termination and validity, it must be that, in equilibrium,
sufficiently many rational players find it in their own interest to check the
validity of the block and to send messages in support of valid blocks. The
problem is that some players might be tempted to free-ride, and let the others
bear the cost of checking. To avoid this situation, it must be that (at least
some) rational players anticipate they are pivotal, i.e., if they fail to
check block validity and send messages in support of valid blocks, this may
derail the process at their own expense.
		% !TeX root = main.tex

To make this point, we look for an equilibrium in which some rational players
check the validity of the block and send a message if and only the block is
valid, and this results in valid blocks being immediately accepted and invalid
blocks being rejected. Before proving that such an equilibrium exists, we
characterise the expected continuation payoff to which it would give rise. 
%To
%do so, we define property $P$ as follows:

%\bigskip

%\bigskip

%\subparagraph{Proposition 3:}
\begin{lemma}
\label{prop3}
\textit{Consider a candidate equilibrium in which some rational players check
the validity of the block and send a message if and only the block is valid,
while the other rational players send messages without checking validity, and
this results in valid blocks being immediately accepted and invalid blocks
being rejected. In such an equilibrium, if it exists, the expected
continuation payoff, at round }$t$\textit{, of the rational players who are to
check block validity is}%
\[
\pi_{check}(t)=R-c_{send}-\phi(t)c_{check},
\]
\textit{while} \textit{the expected continuation payoff, at round }%
$t$\textit{, of the rational players who are not to check block validity is}%
\[
\pi_{send}(t)=R-\psi(t)c_{send},
\]
\textit{where }$\phi(f)=1$, \textit{\ }$\psi(f+1)=1$\textit{\ and both }$\phi
$\textit{\ and }$\psi$\textit{\ satisfy property }$P$ \textit{defined below.}
\end{lemma}
%\bigskip

%\subparagraph{Definition:}
\begin{definition}
\textit{A function }$g$\textit{\ satisfies property }$P$\textit{, if
}$g(t)=1+\frac{f-t+1}{n-t+1}g(t+1),\forall t<f$.
\end{definition}

%\subparagraph{Intuition for Proposition 3:}
%The intuition of the Lemma \ref{prop3} is as follows.
In the candidate equilibrium, participants will reach a point at which the
block is valid and all rational players send a message so that the block
accepted. This gives rise to a payoff $R-c_{send}$, the first part of
$\pi_{check}(t)$. The second part of $\pi_{check}(t)$, $\phi(t)c_{check}$, is
the expected cost of checking block validity, where $\phi(t)$ is the expected
number of times the player expects to check validity before a block is
accepted. Similarly, in $\pi_{send}(t)$, $\psi(t)c_{check}$, is the expected
cost of sending messages, where $\psi(t)$ is the expected number of times the
player expects to send messages before a block is accepted.

%\bigskip
		% !TeX root = main.tex

\begin{proofL}
	We prove this Lemma in 2 parts:
	\begin{enumerate}
		\item Proof of the first part of the proposition, concerning the rational players
		who are expected to check validity:
		
		At round $t=f$, players know that all $f-1$ previous proposers were Byzantine
		and that there are now $n-f+1$ potential proposers, out of which only one is
		Byzantine and $n-f>\nu$ are rational. The expected gains of the rational
		players who are supposed to check are%
		\[
		-c_{check}+\frac{n-f}{n-f+1}(R-c_{send})+\frac{1}{n-f+1}(R-c_{send}),
		\]
		where the first term is the cost of checking validity, the second term
		corresponds to the case in which the current proposer is rational and proposes
		a valid block that is immediately accepted, and the third term corresponds to
		the case in which the proposer is Byzantine, the block is rejected, and we
		move to the next round, at which a valid block is finally accepted (without
		needing any further validity check). This equilibrium payoff simplifies to%
		\[
		R-c_{send}-c_{check},
		\]
		reflecting that eventually a valid block will be accepted, and that from round
		$f$ on the player will need to check validity only once. This equilibrium
		payoff implies that%
		\[
		\phi(f)=1.
		\]

		Now turn to round $t<f$. If round $t\leq f$ is reached, the previous $t-1$
		proposers were Byzantine. There remains $n-(t-1)$ potential proposers. Out of
		them a fraction%
		\[
		\frac{f-(t-1)}{n-t+1}%
		\]
		is Byzantine, while the complementary fraction%
		\[
		\frac{n-f}{n-t+1}%
		\]
		is rational. This fraction being the probability that the next proposer is rational.
		
		To prove the property stated in the Proposition by backward induction, we now
		prove that if this property is satisfied at round $t+1$, that is if%
		\[
		\pi_{check}(t+1)=R-c_{send}-\phi(t+1)c_{check},
		\]
		then it is satisfied at round $t$.
		
		Suppose the rational player follows the equilibrium strategy of checking and
		sending iff the block is valid. Its expected gain from round $t$ on is%
		\[
		-c_{check}+\frac{n-f}{n-t+1}(R-c_{send})+\frac{f-(t-1)}{n-t+1}\pi(t+1),
		\]
		where the first term is the cost of checking the block at round $t$, the
		second term is the probability that the block is valid and accepted multiplied
		by the payoff in that case, and the third term is the probability that the
		block is invalid and rejected multiplied by the payoff in that case.
		Substituting the value of $\pi_{check}(t+1)$, using that the property is
		verified at round $t+1$, the expected gain writes as%
		\[
		-c_{check}+\frac{n-f}{n-t+1}(R-c_{send})+\frac{f-(t-1)}{n-t+1}(R-c_{send}%
		-\phi(t+1)c_{check}).
		\]
		That is%
		\[
		R-c_{send}-\left(  1+\frac{f-(t-1)}{n-t+1}\phi(t+1)\right)  c_{check},
		\]
		which, using the definition of $\phi(t)$, is $R-c_{send}-\phi(t)c_{check}$.
		
		\item Proof of the second part of the proposition, concerning the rational
		players who are just expected to send messages:
		
		Again, we prove that if the property is satisfied at round $t+1$, i.e.,
		$\pi_{send}(t+1)=R-\psi(t+1)c_{send}$, then it is satisfied at round $t$.
		Suppose the rational player follows the equilibrium strategy of not checking
		blocks' validity and always sending a message. Its expected gain from round
		$t$ on is%
		\[
		c_{send}+\frac{n-f}{n-t+1}R+\frac{f-t+1}{n-t+1}\pi_{send}(t+1),
		\]
		where the first term is the cost of sending a message at round $t$, the second
		term is the probability that the block is valid and accepted multiplied by the
		payoff in that case, and the third term is the probability that the block is
		invalid and rejected multiplied by the payoff in that case. Substituting the
		value of $\pi_{send}(t+1)$, the expected gain writes as%
		\[
		-c_{send}+\frac{n-f}{n-t+1}R+\frac{f-t+1}{n-t+1}(R-\psi(t+1)c_{send}).
		\]
		That is%
		\[
		R-\left(  1+\frac{f-t+1}{n-t+1}\psi(t+1)\right)  c_{send},
		\]
		which, using the definition of $\psi(t)$, is $R-\psi(t)c_{send}$.
	\end{enumerate}
	
%QED
\renewcommand{\toto}{prop3} 
\end{proofL}

Relying on  Lemma \ref{prop3}, we now establish that our candidate equilibrium is
indeed an equilibrium. To do so denote the highest index of all Byzantine
players by $i_{B}$.
		% !TeX root = main.tex

%\bigskip

%\subparagraph{Proposition 4:}
\begin{proposition}
\label{prop4}
\textit{When }$f<\nu$\textit{\ and }$n-f>\nu$\textit{, if the cost }$\kappa$
\textit{of accepting an invalid block is large enough, in the sense that}%
\[
\kappa>\alpha(t)c_{check}-\beta(t)c_{send},\forall t<f,
\]
\textit{where}%
\[
\alpha(t)=\frac{(n-t+1)\phi(t)-(f-t+1)\Pr(i_{B}\geq n-\nu+f+2|T\geq
t)\phi(t+1)}{(f-t+1)\Pr(i_{B}<n-\nu+f+2|T\geq t)}%
\]
\textit{and}%
\[
\beta(t)=\frac{\Pr(i_{B}\geq n-\nu+f+2|T\geq t)}{\Pr(i_{B}<n-\nu+f+2|T\geq
t)},
\]
\textit{and if the reward is large enough relative to the costs in the sense
that}%
\[
R\geq\max\left[  \frac{n}{n-f}c_{send},c_{send}+\frac{n}{n-f}c_{check}\right]
,
\]
\textit{\ there exists a Perfect Bayesian Nash equilibrium in which the
strategy of rational players is the following:}

\begin{itemize}
\item \textit{As proposer, a rational player proposes a valid block.}

\item \textit{At any round }$t\leq f$, \textit{when receiving a proposed
block, (i) the rational players with index }$i\in\{t,\dots ,$\textit{\ }%
$n-\nu+f+1\}$\textit{\ check the block validity and send a message only if the
block is valid, while (ii) the rational players with index }$i\in
\{n-\nu+f+2, \dots, n\}$\textit{\ do not check the validity of the block but send a
message. }

\item \textit{If round }$t=f+1$\textit{\ is reached, rational players send a
message without checking if the block is valid. At this point the block is
valid and accepted. }
\end{itemize}

\textit{Hence, in equilibrium, termination occurs no later than at round
}$f+1$.
\end{proposition}

%\bigskip

%\subparagraph{Intuition for Proposition 4:}
%The intuition of Proposition \ref{prop4} is as follows.
On the equilibrium path, invalid blocks (proposed by Byzantine players) are
rejected, while valid blocks (proposed by rational players) are accepted. This
implies that, if round $t=f+1$ is reached, the players know that during all
the previous ($f$) rounds the proposers were Byzantine (to draw this
inference, the rational players use their anticipation that all participants
play equilibrium strategies; hence the Perfect Bayesian nature of the
equilibrium). Consequently, at round $f+1$, the proposer must be rational, and
all players anticipate the proposed block is valid. So, no rational player
needs to check the validity of the block but all send a message, which brings
them expected gain equal to $R-c_{send}$. This is larger than their gain from
deviating (e.g., by not sending a message or by checking the block.)

At previous rounds $t\leq f$, players know that all $t-1$ previous proposers
were Byzantine and that there remains $f-t+1$ Byzantine players with index
strictly larger than $t-1$ (as above, this rational inference is a feature of
the Perfect Bayesian equilibrium we characterize). Do the equilibrium
strategies of the rational players preclude acceptance of an invalid block by
Byzantine processes? To examine this point, consider the maximum possible
number of messages that can be sent if the proposer is Byzantine. {}In
equilibrium the $\nu-f-1$ players with indexes strictly larger than
$n-\nu+f+1$ are to send a message without checking it.\ The worse case
scenario (maximizing the number of messages sent when the block is invalid) is
that none of these players are Byzantine. In that case, in equilibrium, the
number of messages sent when the block is invalid is $f+(\nu-f-1)=\nu-1$, so
that we narrowly escape validation of the invalid block. In contrast, if one
of the rational players deviated from equilibrium and sent a message without
checking the block, in the worse case scenario, this would lead to accepting
an invalid block. Thus, in that sense, the rational players with index
strictly lower than $n-\nu+f+1$ are pivotal. Hence they check block validity,
because, under the condition stated in the proposition, the cost of accepting
an invalid block is so large that rational players do not want to run that risk.\ 
		% !TeX root = main.tex

\begin{proofP}
For clarity, we decompose the proof in 5 steps.
\begin{enumerate}
	\item The first step is to note that rational proposers strictly prefer to
	propose a valid block than an invalid one. This is because, when they follow
	their equilibrium strategy of proposing a valid block, it is accepted and the
	proposer gets $R-c_{check}-c_{send},$ while if they propose an invalid block,
	it is rejected, and we move to the next round, a which, in equilibrium, the
	player gets at most $R-c_{check}-c_{send}$ (and possibly less). Indeed, this
	player incurs the cost of checking validity at the next round, because the
	rational players who are not expected to check validity have indexes above
	$n-\nu+f+1$, which are above $f+1$, so that they do not get to propose blocks.
	
	\item The next step concerns the actions of the rational players when round
	$t=f+1$ is reached. At that round, all players know the proposer must be
	rational and the proposed block valid. In equilibrium no rational checks
	validity but all send a message. Any other action would be dominated.
	
	\item\label{mainDeviation} The third step concerns the most relevant deviation, in which a rational
	player expected to check block validity fails to do so. If at round $t$ a
	rational player supposed to check, deviates and sends a message without
	checking block validity, its expected continuation payoff is%
	\begin{align*}
	\left(  1-\frac{f-(t-1)}{n-t+1}\right)  \left(  R-c_{send}\right)
	+\frac{f-(t-1)}{n-t+1}\Pr(i_{B}  & <n-\nu+f+1)\left(  R-c_{send}-\kappa\right)
	\\
	+\frac{f-(t-1)}{n-t+1}\Pr(i_{B}  & \geq n-\nu+f+1)\left(  \pi(t+1)-c_{send}%
	\right).
	\end{align*}
	The first term is the payoff obtained by the deviating rational player if the
	current block is valid, and therefore immediately accepted. The second term is
	the payoff obtained by the deviating player when he was pivotal and triggered
	acceptance of an invalid block. To see this, consider the number of messages
	when the block is invalid, the rational player is deviating and the indexes of
	all the Byzantine players are strictly lower than $n-\nu+f+2$: $f$ messages
	are sent by the Byzantine processes, 1 message is sent by the deviating
	rational agent, $\nu-f-1$ messages are sent by the rational players with index
	above than or equal to $n-\nu+f+2$. The resulting total number of messages is
	$\nu$ and the block is accepted. The last term corresponds to the case in
	which the deviating rational player is not pivotal, and the invalid block is
	not accepted, so that we move to the next round.
	
	Substituting the value of $\pi_{check}(t+1)=R-c_{send}-\phi(t+1)c_{check}$
	from Lemma \ref{prop3}, the expected continuation value of the deviating player is%
	
	\begin{align*}
	\left(  1-\frac{f-(t-1)}{n-t+1}\right)  \left(  R-c_{send}\right)
	&+\frac{f-(t-1)}{n-t+1}\Pr(i_{B}   <n-\nu+f+2|T\geq t)\left(  R-c_{send}%
	-\kappa\right) \\
	+\frac{f-(t-1)}{n-t+1}\Pr(i_{B}  & \geq n-\nu+f+2|T\geq t)\left(
	R-c_{send}-\phi(t+1)c_{check}-c_{send}\right).
	\end{align*}
	Or%
	\begin{align*}
	\left(  R-c_{send}\right)  -\frac{f-(t-1)}{n-t+1}\Pr(i_{B}  & <n-\nu+f+2|T\geq
	t)\kappa\\
	-\frac{f-(t-1)}{n-t+1}\Pr(i_{B}  & \geq n-\nu+f+2|T\geq t)\left(
	\phi(t+1)c_{check}+c_{send}\right).
	\end{align*}
	The equilibrium condition is that this deviation payoff must be lower than the
	equilibrium continuation payoff of the player%
	
	\[
	R-c_{send}-\phi(t)c_{check}.
	\]
	That is%
	\begin{align*}
	\frac{f-(t-1)}{n-t+1}\Pr(i_{B}  & <n-\nu+f+2|T\geq t)\kappa>\phi(t)c_{check}\\
	-\frac{f-(t-1)}{n-t+1}\Pr(i_{B}  & \geq n-\nu+f+2|T\geq t)\left(
	\phi(t+1)c_{check}+c_{send}\right).
	\end{align*}

	Note that
	\[
	\phi(t)\geq\frac{f-(t-1)}{n-t+1}\Pr(i_{B}\geq n-\nu+f+2|T\geq t)\phi(t+1),
	\]
	since by the definition of $\phi(t)$ this inequality is equivalent to
	\[
	1+\frac{f-(t-1)}{n-t+1}\phi(t+1)\geq\frac{f-(t-1)}{n-t+1}\Pr(i_{B}\geq
	n-\nu+f+2|T\geq t)\phi(t+1),
	\]
	which indeed holds. Thus we can write the equilibrium condition as%
	\[
	\kappa>\alpha(t)c_{check}-\beta(t)c_{send},\forall t<f,
	\]
	as stated in the proposition.
	
	\item Other possible deviations for rational player supposed to check block's
	validity are easier to rule out:
	
	First, the player could do nothing (neither check nor send).\ Relative to the
	equilibrium payoff, this deviation economises the cost of checking
	($c_{check}$). If the current proposer is Byzantine, the player then obtains
	the same payoff after a one shot deviation as on the equilibrium path
	($\pi_{check}(t+1)$). If the current proposer is rational, the block gets
	accepted, but the player does not earn any reward. So the deviation is
	dominated if%
	\[
	\frac{n-f}{n-t+1}(R-c_{send})\geq c_{check},
	\]
	which holds under the condition, stated in the proposition, that $R\geq
	\max\left[  \frac{n}{n-f}c_{send},c_{send}+\frac{n}{n-f}c_{check}\right]  $.
	
	Second, the player could check the block validity, and then send a message
	irrespective of whether the block is valid or not. This would generate a lower
	payoff than the main deviation, shown above (in \ref{mainDeviation}.) to be dominated.
	
	Third, the player could check validity but then send no message. When the
	current proposer is Byzantine, this one-shot deviation yields the same payoff
	as the equilibrium strategy. When the current proposer is rational, this
	deviation yields a payoff of $-c_{check}$, which is lower than the equilibrium
	payoff $R-c_{send}-c_{check}$.
	
	Fourth, the player could check the block's validity and send a message only if
	the block is invalid, which is trivially dominated.
	
	\item Finally turn to deviations of rational players supposed to send messages
	without checking blocks' validity.
	
	First, consider the possibility to abstain from sending a message. This
	economises the costs $c_{send}$, but, in case the block is valid and accepted,
	this implies the agent loses the reward $R$. So, the deviation is dominated if%
	\[
	\frac{n-f}{n-t+1}R\geq c_{send},
	\]
	which holds under the condition, stated in the proposition, that $R\geq
	\max\left[  \frac{n}{n-f}c_{send},c_{send}+\frac{n}{n-f}c_{check}\right]  $.
	
	Second, consider the possibility of checking validity and sending a message
	only for valid blocks. This deviation would imply the agent would have to incur
	the cost of checking ($c_{check}$), but it would economise the cost of sending
	a message when the block is invalid. So the deviation is dominated if%
	\[
	c_{check}\geq\frac{f-t+1}{n-t+1}c_{send},
	\]
	which holds because of our assumption that $c_{check}\geq c_{send}$.
	
	Other deviations, such as checking validity but never sending messages, or
	checking validity and always sending messages, or checking validity and
	sending only if the block is invalid, are trivially dominated.
\end{enumerate}
\renewcommand{\toto}{prop4} 
\end{proofP}

\section{Conclusion and Future Work} \label{sec:conclusion}
%\input{conclusion}

%\section{Next steps}\label{sec:nextStep}
	% !TeX root = main.tex

In this paper we model PBFT-consensus based blockchains as a coordination game between rational and Byzantine processes. We derive the conditions (on the majority threshold and the proportion of Byzantine processes) under which consensus properties are guaranteed in equilibrium or not. In future work, we will extend the analysis to more general Byzantine strategies and rational agents preferences, costs and rewards.

%%%%%%%%version avec les idées
%In this paper we model Byzantine Fault-Tolerant blockchains as a game between rational and Byzantine processes. We derived the conditions (on the threshold $\nu$ and proportion of Byzantine processes,$f$) under which rational players reach an equilibrium where the consensus properties are guaranteed. In future work we will endogenise the behaviour of the Byzantine players assuming they choose their strategies to minimise the expected gains of the rational players. Moreover, while in the curent paper we analyse equilibrium in the protocol when players receive the reward R only when they sent a message, in future work we will analyse equilibrium when all processes receive the reward R (irrespective of whether they sent a message or not.) We will thus study which of the two reward mechanisms gives rise to the best equilibrium outcome. Finally, while in the current paper rational players are patient, in future work we will assume they have discount factor $\beta<1$. The idea is that rational players are better off if consensus is reached early than if it is delayed.   
% 

\newpage
\bibliographystyle{plain}
\bibliography{biblio}
%\newpage
%\appendix
%\section*{Appendix}
%	\input{algoVS}
	%\input{decisionTree}
%	\input{proofs}
%	\input{protocols}
\end{document}